# Epitaxy of Advanced Nanowire Quantum Devices

Sasa Gazibegovic,[1,2,*] Diana Car,[1,2,*] Hao Zhang,[1,*] Stijn C. Balk,[1] John A. Logan,[3] Michiel W. A. de Moor,[1] Maja C. Cassidy,[1] Rudi Schmits,[4] Di Xu,[1] Guanzhong Wang,[1] Peter Krogstrup,[5] Roy L. M. Op het Veld,[1,2] Jie Shen,[1] Daniël Bouman,[1] Borzoyeh Shojaei,[3] Daniel Pennachio,[3] Joon Sue Lee,[6] Petrus J. van Veldhoven,[2] Sebastian Koelling,[2] Marcel A. Verheijen,[2,7] Leo P. Kouwenhoven,[1,8] Chris J. Palmstrøm,[3,6,9] Erik P.A.M. Bakkers[1,2]

*[1] QuTech and Kavli Institute of NanoScience, Delft University of Technology, 2600 GA Delft, the Netherlands*
*[2] Department of Applied Physics, Eindhoven University of Technology, 5600 MB Eindhoven, the Netherlands*
*[3] Materials Department, University of California, Santa Barbara, CA 93106, United States*
*[4] TNO Technical Sciences, Nano-Instrumentation Department, 2600 AD Delft, the Netherlands*
*[5] Center for Quantum Devices and Station-Q Copenhagen, Niels Bohr Institute, University of Copenhagen, 2100 Copenhagen, Denmark*
*[6] California NanoSystems Institute, University of California, Santa Barbara, CA 93106, United States*
*[7] Philips Innovation Services Eindhoven, High Tech Campus 11, 5656AE Eindhoven, the Netherlands*
*[8] Microsoft Station-Q at Delft University of Technology, 2600 GA Delft, the Netherlands*
*[9] Electrical and Computer Engineering, University of California, Santa Barbara, CA 93106, United States*

**Semiconductor nanowires provide an ideal platform for various low-dimensional quantum devices. In particular, topological phases of matter hosting non-Abelian quasi-particles can emerge when a semiconductor nanowire with strong spin-orbit coupling is brought in contact with a superconductor[1,2]. To fully exploit the potential of non-Abelian anyons for topological quantum computing, they need to be exchanged in a well-controlled braiding operation[3-8]. Essential hardware for braiding is a network of single-crystalline nanowires coupled to superconducting islands. Here, we demonstrate a technique for generic bottom-up synthesis of complex quantum devices with a special focus on nanowire networks having a predefined number of superconducting islands. Structural analysis confirms the high crystalline quality of the nanowire junctions, as well as an epitaxial superconductor-semiconductor interface. Quantum transport measurements of nanowire "hashtags" reveal Aharonov-Bohm and weak-antilocalization effects, indicating a phase coherent system with strong spin-orbit coupling. In addition, a**

**proximity-induced hard superconducting gap is demonstrated in these hybrid superconductor-semiconductor nanowires, highlighting the successful materials development necessary for a first braiding experiment. Our approach opens new avenues for the realization of epitaxial 3-dimensional quantum device architectures.**

Majorana Zero Modes (MZMs) are predicted to emerge once a superconductor (SC) is coupled to a semiconductor nanowire (NW) with a strong spin-orbit interaction (SOI) in an external magnetic field[1,2]. InSb NWs are a prime choice for this application due to the large Landé g-factor (~50) and strong Rashba SOI[9], crucial for realization of MZMs. In addition, InSb nanowires generally show high mobility and ballistic transport[10-12]. Indeed, signatures of Majorana zero modes (MZMs) have been detected in hybrid superconductor-semiconductor InSb and InAs NW systems[11,13-15]. Multiple schemes for topological quantum computing based on braiding of MZMs have been reported, all employing hybrid NW networks[3-8].

Top-down fabrication of InSb NW networks is an attractive route towards scalability[16], however, the large lattice mismatch between InSb and insulating growth substrates limits the crystal quality. An alternative approach is bottom-up synthesis of out-of-plane NW networks which, due to their large surface-to-volume ratio, effectively relieve strain on their sidewalls, enabling the growth of single-crystalline NWs on highly lattice-mismatched substrates[17-19]. Recently, different schemes have been reported for merging NWs into networks[20-22]. Unfortunately, these structures are either not single-crystalline, due to a mismatch of the crystal structure of the wires with that of the substrate (*i.e.* hexagonal NWs on a cubic substrate)[22], or the yield is low due to the limited control over the multiple accessible growth directions (the yield decreases with the number of junctions in the network)[23].

Here, we develop a technique for bottom-up synthesis of single-crystalline InSb NW networks with an unprecedented yield of crossed junctions. Accurate control over the NW position and growth direction enables us to grow complicated networks of up to four crossed junctions, such as closed loops of four interconnected nanowires (referred to as "hashtags"). Furthermore, this platform allows *in-situ* growth of a predefined number of separated superconducting islands (SCIs) on the NWs. This eliminates the need for metal etching during device fabrication. Therefore, the integration of semiconductors with metals (*e.g.* niobium) is possible without an additional etching process. This guarantees that the pristine atomically flat InSb (110) facets are left intact, a key-element for high device performance. At the same time a clean epitaxial superconductor-NW interface is established, which has recently been proven to be crucial for the quality of the induced superconducting gap[24,25].

For the growth of the nanowire networks a substrate with trenches is first fabricated (see Fig. 1a). These structures are defined by e-beam lithography (EBL), a reactive ion etch and a subsequent wet etch to expose (111)B facets on an InP (100) crystal surface (Fig. 1a). A second lithography step is then used to position gold particles, which catalyse NW growth via the vapor-liquid-solid (VLS) mechanism, on the inclined facets. Due to the geometry of the (111)B facets, the NWs are forced to grow towards each other and can fuse into a network. The final size and symmetry of the networks are controlled by the dimensions of the trenches and the spacing between them, *i.e.* the parameters *a*-*e*, as indicated in Fig. 1a. The *left* (*right*) trenches and wires grown from them are labelled $L_1$, $L_2$ ($R_1$, $R_2$). The offset ($\Delta y$) between the gold particles is an important parameter to control because: first, if $\Delta y < D$, where $D$ is the NW diameter, NWs will merge during growth (specifically, for $\Delta y \sim 0$ ($\Delta y \leq D$), resulting in the formation of a T-junction (X-junction)[23]); second, $\Delta y > D$ enables shadow-growth of the SCIs, as discussed later in the text. Fig. 1b shows a uniform array of InP NW which are used as stems to facilitate uniform nucleation of InSb NWs. When InSb NWs (highlighted in red in Fig. 1c-f) are grown on top of these InP stems, NW networks with 1-4 wire-wire junctions are formed, depending on the trench design (Fig. 1c-f). Importantly, this approach is generic and can be used to synthesize interconnected nanowires of various semiconductor materials which grow along a <111>B direction. The number of wire-wire junctions can be increased by allowing for longer NW growth times and/or fabricating a larger number of *left* and *right* trenches. High crystalline quality of the InSb nanowire junctions is confirmed by high-resolution transmission electron microscopy (HRTEM) imaging of a "hashtag" structure (see Extended Data Fig. 6).

Next, we combine the NW-network geometry with the directionality of molecular beam epitaxy (MBE) to shadow-grow aluminium SCIs on the InSb wires. The aluminium flux is aligned parallel to the trenches (Fig. 2a), such that a frontal wire casts a shadow on a wire in the background (Inset Fig. 2a). This causes interruptions in a uniform layer of aluminium as shown in Fig. 2b (left). For an effective shadowing, it is important that the frontal wire does not merge with the shadowed wire, *i.e.* $\Delta y > D$. The number of shadows, $n$, (and, accordingly, the number of SCIs, $n+1$) on any InSb NW is determined by the number of wires directly in front of that NW. For example, Fig. 2b (right) depicts an InSb NW with three shadows cast by three frontal NWs. The position and the width of the shadows are uniform for all wires examined and are set by the relative position of the wires and the solid angle of the aluminium effusion cell. The abrupt transition between the shadowed region of the NW and the segment covered with aluminium is evident from the chemical composition map (Fig. 2c) acquired by energy-dispersive X-ray spectroscopy (EDX) combined with scanning transmission electron microscopy (STEM). The line-of-sight directionality of MBE growth results in aluminium being deposited on two out of six facets of an InSb NW, as can be seen from a STEM-EDX map of a nanowire cross-section (Fig. 2d). The partial coverage of a NW with aluminium is essential as it allows tuning of the electron density of the proximitised NW by an external gate electrode, which is necessary for accessing the topological phase. The epitaxial interface between the InSb NW and a uniform, thin aluminium layer is revealed by HRTEM imaging (Fig. 2e). In the next section, we assess the electronic quality of our structures.

Phase coherent transport, a basic requirement for braiding, can be verified by the Aharonov-Bohm (AB) effect, induced by coherent interference of electron waves[26]. To investigate the AB effect in our NW networks, NW "hashtags" were transferred onto a $SiO_2$/p-Si substrate and contacted by metal electrodes (Au/Cr, Fig. 3a left inset). Fig. 3a shows the magnetoconductance of a representative device (Device A). Periodic AB oscillations can be clearly seen (Fig. 3a right inset), as well as a pronounced weak-antilocalization (WAL) conductance peak at $B$ = 0 T. The WAL conductance peak is present in most of the measured "hashtag" devices, for both in- and out-of-plane magnetic field orientations (see Extended Data Fig. 7), indicating a strong spin-orbit coupling in this system. The period of the AB oscillations is extracted from a discrete Fourier transform of the magnetoconductance. Figure 3b shows the averaged FFT spectrum whose peak frequency (60 ± 2 $T^{-1}$) corresponds to a period $\Delta B$ of 16.7 ± 0.6 mT. The effective area ($A$) calculated from this AB period ($A = \Phi/\Delta B$ = 0.25 ± 0.01 μm$^2$, where $\Phi = h/e$ is the flux quantum) is in agreement with the measured area of a "hashtag" loop ($A$ ~ 0.25 ± 0.02 μm$^2$). We determine the peak frequency for four different devices with different loop areas, showing good agreement with the expected values (Fig. 3b inset). This agreement between the theory and the experiment reveals that the observed AB oscillations are indeed due to the quantum interference of electron waves emanating from the two transport channels that constitute the "hashtag".

Magnetoconductance traces taken at increasing temperature values are shown in Fig. 3c. AB oscillations persist up to ~1.6 K. The amplitude of the AB oscillations decays exponentially with temperature (Fig. 3d). This indicates that the phase coherence length is proportional to $T^1$, as expected for decoherence due to dephasing in a quasiballistic system coupled to a thermal bath[27]. From the slope, we can estimate a phase coherence length of 0.7 ± 0.1 μm at 1 K, which translates to 2.3 ± 0.3 μm at 300 mK.

The last essential ingredient for a topological phase is induced superconductivity in the InSb NWs. For this study InSb wires with two SCIs were used to fabricate N-NW-S devices by replacing one SCI with a normal metal electrode (see Supplementary Note 2.2). The shadowed region of the nanowire is situated in between the normal contact and the other SCI, and can be depleted by a bottom gate (inset Fig. 4c), to form a tunnel barrier. In the tunnelling regime, the differential conductance reflects the quasiparticle density-of-states in the proximitised nanowire segment. Fig. 4a shows a plot of differential conductance ($dI/dV$) vs. bias voltage ($V$) and back-gate voltage ($V_{gate}$) at 20 mK. Hence, the two high-conductance horizontal lines (at $V = \pm 0.24$ mV) in Fig. 4a correspond to the superconducting coherence peaks. The shape of the superconducting gap can be clearly resolved in Fig. 4b, which shows a vertical line-cut plotted on both linear (left) and logarithmic (right) scale, indicating the ratio of the above-gap to sub-gap conductance $G_N/G_S \sim 100$, comparable to the value reported for InAs NWs fully surrounded by an aluminium shell[25]. Fig. 4c maps out the obtained values of $G_S$ vs. $G_N$ (black dots) together with the Beenakker expression (red line) for an N-QPC-S system. This expression assumes that $G_S$ is due to a single channel Andreev reflection in the shadowed region (see Extended Data Fig. 8)[28]. Theory and experiment are in agreement over two orders of magnitude in conductance. This shows that the $G_S$ in this system is dominated by the Andreev process in the absence of quasi-particle transport. Fig. 4d shows the differential conductance ($dI/dV$) of the same device as a function of bias voltage ($V$) and magnetic field ($B$) pointing along the nanowire, taken at $V_{gate} = -5.7$ V. From the horizontal line-cut at $V = 0$ V (lower panel), it can be clearly seen that $G_S$ is pinned to extremely low values of conductance for magnetic field values up to 0.9 Tesla. The evolution of the induced superconducting gap in the magnetic field is illustrated in Fig. 4e. The black, green and orange line cuts are taken at $B = 0$, 0.5 and 1 T, respectively. Importantly, the induced hard superconducting gap in Al-InSb nanowires endures up to $B \sim 0.9$ T, which surpasses the value of the magnetic field required for achieving a topological phase transition in InSb ($B$ ~ 0.2 T)[1,2,11]. Extended Data Fig. 8 shows data of additional devices and the corresponding analysis.

The combination of phase coherent transport in a network of nanowires and a hard-superconducting gap in InSb NWs, induced by local superconductor islands, is a significant materials advancement that paves the road for the first Majorana braiding experiments. We emphasize that the platform developed in this work is generic and can be used for many different superconductor-semiconductor combinations, opening opportunities in new quantum devices.

**References**:

1. Lutchyn, R. M. *et al.* Majorana Fermions and a Topological Phase Transition in Semiconductor-Superconductor Heterostructures. *Physical Review Letters* **105**, 077001 (2010).
2. Oreg, Y. *et al.* Helical Liquids and Majorana Bound States in Quantum Wires. *Physical Review Letters* **105**, 177002 (2010).
3. Alicea, J. *et al.* Non-Abelian statistics and topological quantum information processing in 1D wire networks. *Nat Phys* **7**, 412 (2011).
4. Hyart, T. *et al.* Flux-controlled quantum computation with Majorana fermions. *Physical Review B* **88**, 035121 (2013).
5. Aasen, D. *et al.* Milestones Toward Majorana-Based Quantum Computing. *Physical Review X* **6**, 031016 (2016).
6. Plugge, S. *et al.* Majorana box qubits. *New Journal of Physics* **19**, 012001 (2017).
7. Vijay, S. & Fu, L. Teleportation-based quantum information processing with Majorana zero modes. *Physical Review B* **94**, 235446 (2016).
8. Karzig, T. *et al.* Scalable Designs for Quasiparticle-Poisoning-Protected Topological Quantum Computation with Majorana Zero Modes. *ArXiv e-prints* **1610** (2016). Preprint at https://arxiv.org/abs/1610.05289.
9. van Weperen, I. *et al.* Spin-orbit interaction in InSb nanowires. *Physical Review B* **91**, 201413 (2015).
10. Kammhuber, J. *et al.* Conductance Quantization at Zero Magnetic Field in InSb Nanowires. *Nano Letters* **16**, 3482 (2016).
11. Zhang, H. *et al.* Ballistic Majorana nanowire devices. *ArXiv e-prints* **1603** (2016). Preprint at https://arxiv.org/abs/1603.04069.
12. Fadaly, E. M. T. *et al.* Observation of Conductance Quantization in InSb Nanowire Networks. *ArXiv e-prints* **1703** (2017). Preprint at https://arxiv.org/abs/1703.05195.

13. Mourik, V. *et al.* Signatures of Majorana Fermions in Hybrid Superconductor-Semiconductor Nanowire Devices. *Science* **336**, 1003 (2012).

14. Deng, M. T. *et al.* Majorana bound state in a coupled quantum-dot hybrid-nanowire system. *Science* **354**, 1557 (2016).

15. Albrecht, S. M. *et al.* Exponential protection of zero modes in Majorana islands. *Nature* **531**, 206 (2016).

16. Shabani, J. *et al.* Two-dimensional epitaxial superconductor-semiconductor heterostructures: A platform for topological superconducting networks. *Physical Review B* **93**, 155402 (2016).

17. Conesa-Boj, S. *et al.* Gold-Free Ternary III–V Antimonide Nanowire Arrays on Silicon: Twin-Free down to the First Bilayer. *Nano Letters* **14**, 326 (2014).

18. Plissard, S. R. *et al.* From InSb Nanowires to Nanocubes: Looking for the Sweet Spot. *Nano Letters* **12**, 1794 (2012).

19. Caroff, P. *et al.* InSb heterostructure nanowires: MOVPE growth under extreme lattice mismatch. *Nanotechnology* **20**, 495606 (2009).

20. Dalacu, D., Kam, A., Austing, D. G., Poole, P. J. & al., e. Droplet Dynamics in Controlled InAs Nanowire Interconnections. *Nano Letters* **13**, 2676 (2013).

21. Kang, J.-H. *et al.* Crystal Structure and Transport in Merged InAs Nanowires MBE Grown on (001) InAs. *Nano Letters* **13**, 5190 (2013).

22. Rieger, T. *et al.* Crystal phase transformation in self-assembled InAs nanowire junctions on patterned Si substrates. *Nano letters* **16**, 1933 (2016).

23. Car, D. *et al.* Rationally Designed Single-Crystalline Nanowire Networks. *Advanced Materials* **26**, 4875 (2014).

24. Krogstrup, P. *et al.* Epitaxy of semiconductor–superconductor nanowires. *Nat Mater* **14**, 400 (2015).

25. Chang, W. *et al.* Hard gap in epitaxial semiconductor-superconductor nanowires. *Nature nanotechnology* **10**, 232 (2015).

26. Washburn, S. & Webb, R. A. Aharonov-Bohm effect in normal metal quantum coherence and transport. *Advances in Physics* **35**, 375 (1986).

27. Stern, A. *et al.* Phase uncertainty and loss of interference: A general picture. *Physical Review A* **41**, 3436 (1990).

28. Beenakker, C. W. J. Quantum transport in semiconductor-superconductor microjunctions. *Physical Review B* **46**, 12841 (1992).

29. White, L. K. Bilayer taper etching of field oxides and passivation layers. *Journal of The Electrochemical Society* **127**, 2687 (1980).

30. Dalacu, D. *et al.* Selective-area vapour–liquid–solid growth of InP nanowires. *Nanotechnology* **20**, 395602 (2009).

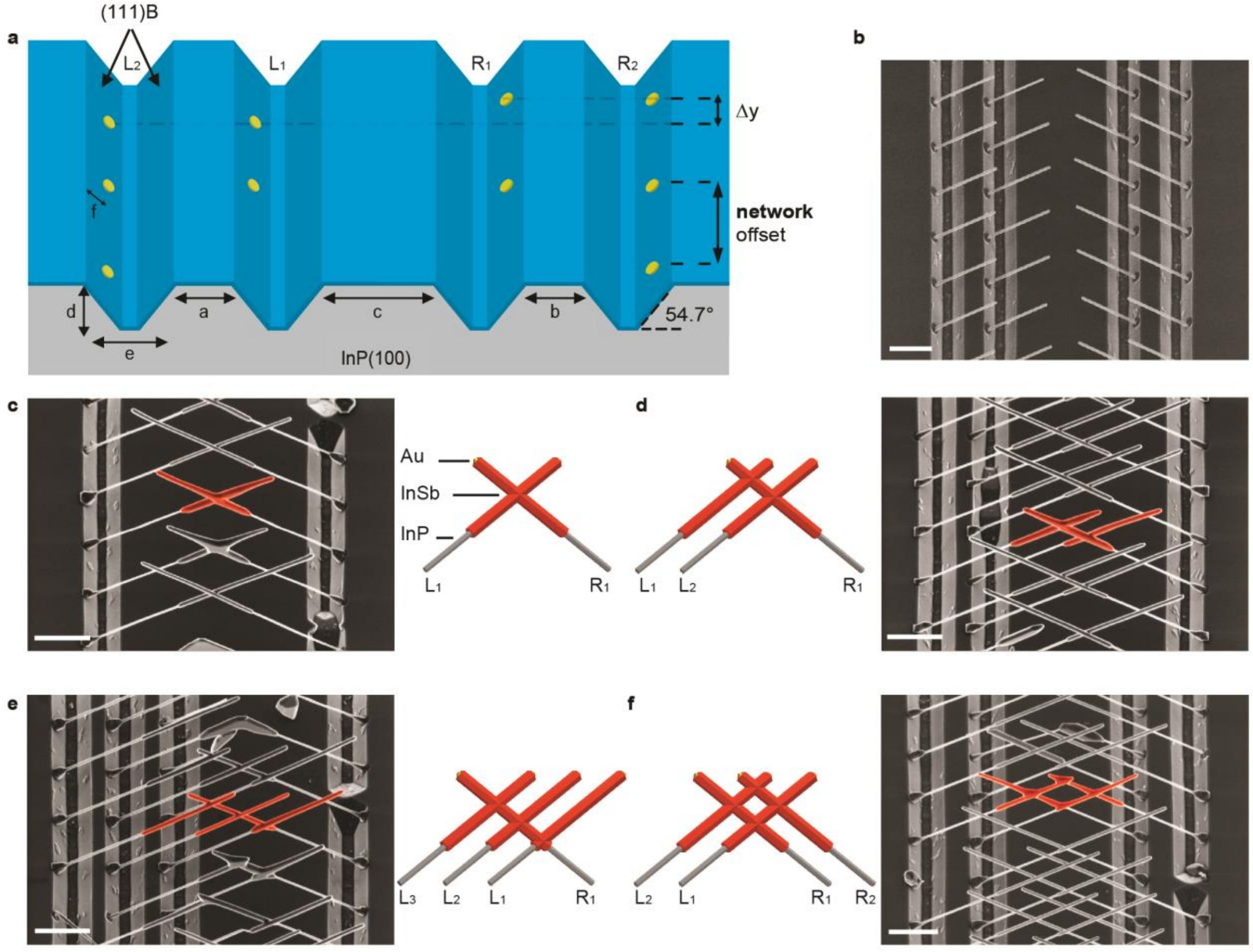

**Figure 1 | Deterministic growth of InSb nanowire networks. a,** Schematic illustration of the substrate with etched trenches. Gold catalysts are lithographically defined on the inclined facets. The offset between the catalyst particles ($\Delta y$) is critical for the realization of nanowire networks and shadowed superconducting islands. The size and the symmetry of the networks are controlled by the dimensions of the trenches indicated in the schematic: the spacing between the *left-left* ($L_1$, $L_2$), *a*, *right-right* ($R_1$, $R_2$), *b*, and *left-right* ($L_1$, $R_1$,) trenches, *c*, as well as the trench depth, *d,* width, *e* and position of the gold particles on the inclined facets, *f*. **b,** A scanning electron microscopy (SEM) image of InP nanowires which serve as stems for InSb nanowire growth. **c-f,** SEM images and schematic illustrations of accomplished NW structures having **c,** 1 junction **d,** 2 junctions **e,** 3 junctions and **f,** 4 junctions (“hashtag”). All SEM images are taken at 30°-tilt. All scale bars are 1 µm.

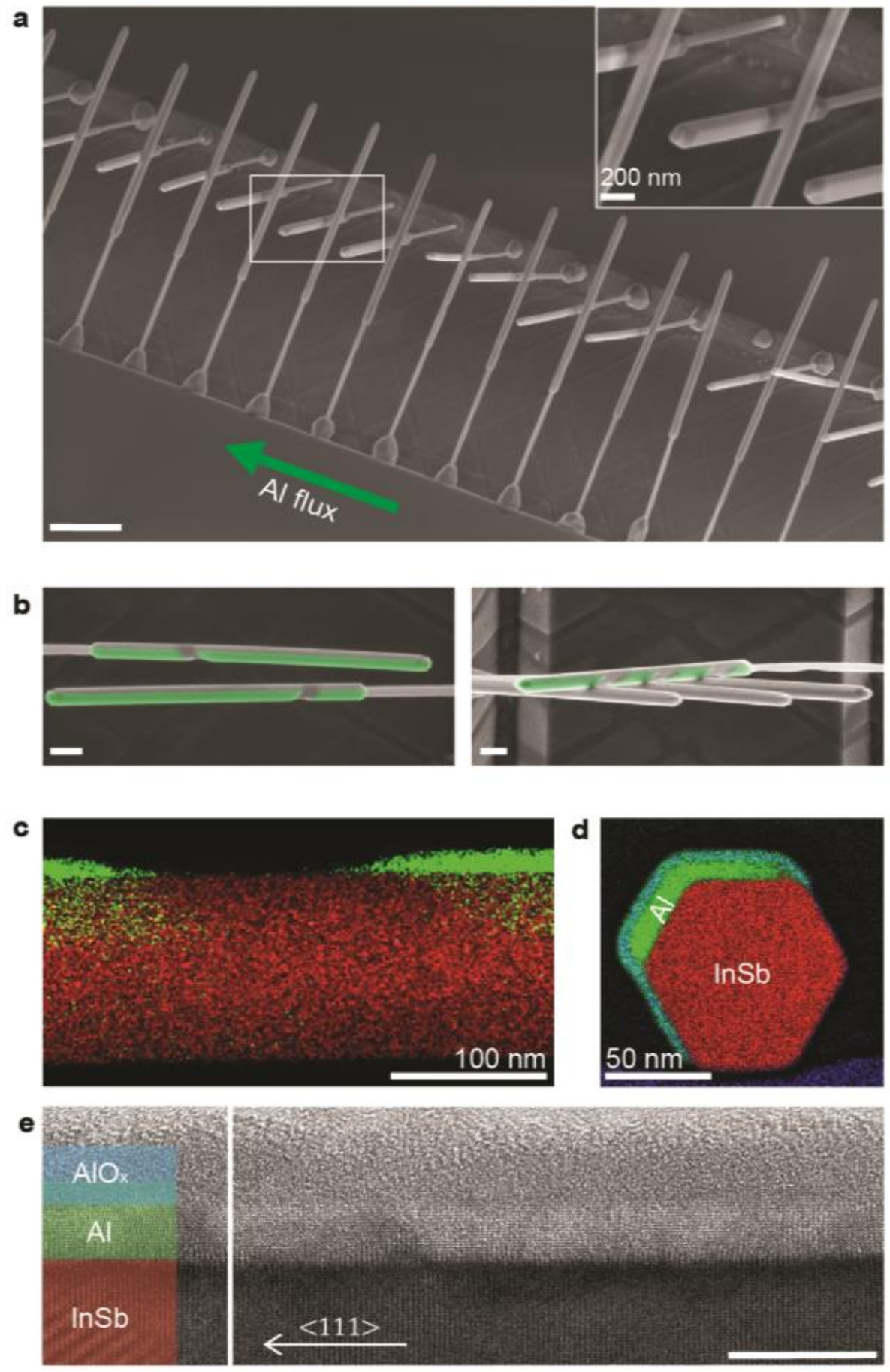

**Figure 2 | Epitaxial growth of Al islands on InSb nanowires. a,** A 45°-tilted SEM image of an array of Al-InSb nanowires. The green arrow indicates the direction of Al beam flux during deposition. Scale bar is 1 μm. Inset: A zoom-in on the area indicated by a white rectangle in the main panel. Each InSb nanowire is covered by two Al islands separated by a shadowed region. The number of shadows, *n*, and hence the number of superconducting islands, *n+1*, is determined by the number of wires directly in front of the shadowed wire. **b,** SEM images of InSb nanowires with two (left) and four (right) Al islands (false-coloured green). Both scale bars are 200 nm. **c,** STEM-EDX chemical composition map of an InSb nanowire (red) with Al islands (green) separated by an Al-free shadowed region. **d,** EDX chemical composition map of the nanowire cross-section. Al (green) is covering two out of the six {110} InSb side-facets. The Al-InSb interface is oxygen-free. **e,** High-resolution transmission electron microscopy (HRTEM) image of an InSb nanowire (red) covered with a thin (~10 nm), crystalline film of Al (green) and a layer of $AlO_x$ (blue). InSb growth direction <111> is indicated by a white arrow. The image is taken along <110> zone axis. The scale bar is 10 nm.

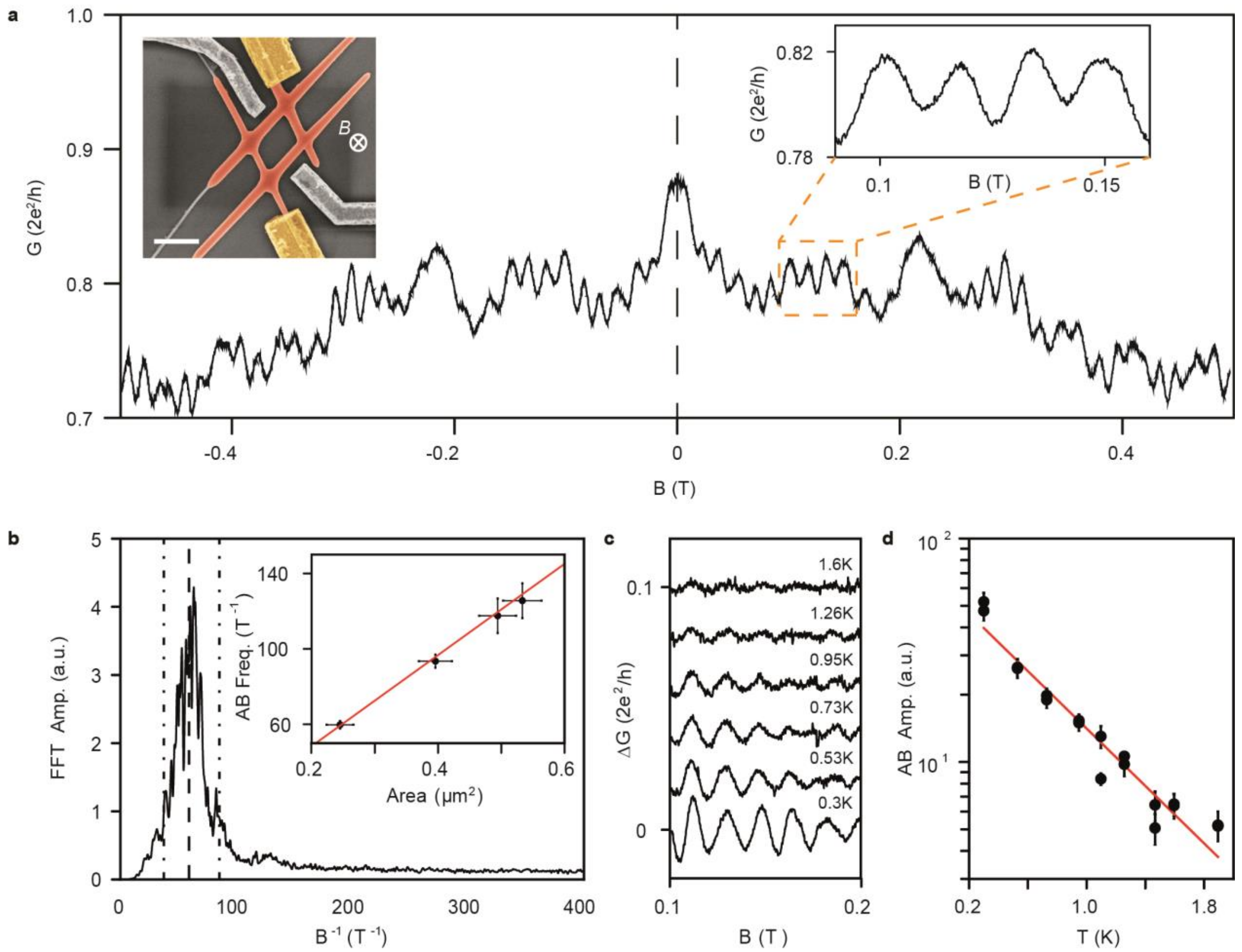

**Figure 3 | Aharonov-Bohm (AB) and weak anti-localization (WAL) effects in nanowire "hashtags". a,** Magnetoconductance of a "hashtag" shows periodic AB oscillations and a WAL peak at $B = 0$ T. **Inset (left)**, A false-coloured SEM image of the device. An InSb "hashtag" (red) is contacted with normal metal electrodes (yellow) and measured in an out-of-plane magnetic field at 300 mK. Scale bar corresponds to 500 nm. **Inset (right)**, A zoom-in on the region indicated by an orange rectangle in the main panel, containing four AB periods. **b,** FFT spectrum of the magnetoconductance of this device (ensemble averaged), indicating the AB oscillation frequency. The dashed line indicates the expected frequency based on the area calculated from the SEM image, dash-dotted lines indicate the expected minimum and maximum frequencies due to the finite thickness of the interferometer arms. **Inset**, Plot of the peak frequency, assigned from the averaged FFT spectra, as a function of the loop area for four different hashtag devices. The red line corresponds to the expected frequency of a $h/e$ periodic oscillation for a given loop area. **c**, Temperature dependence of AB oscillations (background subtracted). The AB effect persists up to 1.6 K. Curves are offset vertically for clarity. **d,** AB amplitude as a function of temperature. The red line is a fit to the data, showing an exponential decay of the oscillation amplitude, as expected for a quasiballistic 1D channel[27].

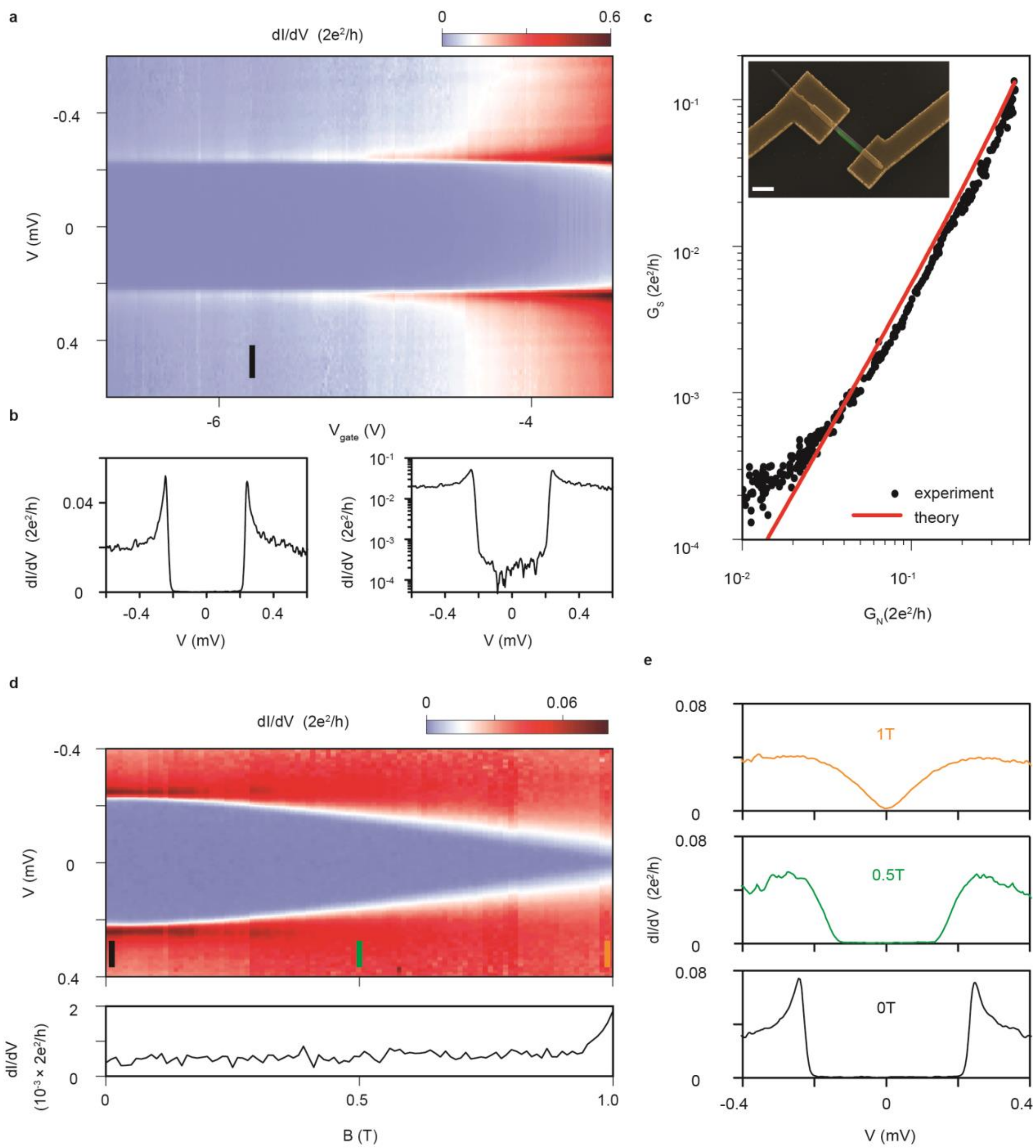

**Figure 4 | Hard induced superconducting gap in shadowed Al-InSb nanowire device. a**, Differential conductance (*dI/dV*) as a function of bias voltage ($V$) and back gate voltage ($V_{gate}$) in the tunneling regime, resolving a hard superconducting gap (at ~20 mK). **b**, A line cut taken at the position indicated by the black bar in panel **a** (at $V_{gate}$ = -5.8 V), plotted on linear (left) and logarithmic scale (right). The ratio of above-gap and sub-gap conductance ($G_N/G_S$) reaches ~100. The induced superconducting gap size is $\Delta$ ~ 0.24 meV. **c**, Sub-gap conductance as a function of above-gap conductance. The red line is the theoretical curve calculated assuming only Andreev processes[28]. **Inset**, A false-coloured SEM image of the similar device. The device is an N (yellow)-NW (grey)-S (green) system. **d**, Magnetic field dependence of the

superconducting gap ($V_{gate}$ ~ -5.7 V) in the device. The magnetic field direction is aligned with the nanowire axis. The lower panel shows a horizontal line-cut taken at $V$ = 0 V (in the middle of the superconducting gap). **e**, Vertical line cuts taken at positions indicated by a black ($B$ = 0 T), green ($B$ = 0.5 T) and orange bar ($B$ = 1 T) in the upper panel **d**, illustrating the evolution of the induced superconducting gap in the increasing magnetic field.

*These authors contributed equally to this work.

**Acknowledgments**

We would like to thank to K. Zuo and Y. Vos for the assistance in fabrication and measurement of Aharonov-Bohm devices. We would like to acknowledge N. Wilson for the assistance at University of Santa Barbara California. This work has been supported by the European Research Council (ERC HELENA 617256 and Synergy), the Dutch Organization for Scientific Research (NWO-VICI 700.10.441), the Foundation for Fundamental Research on Matter (FOM) and Microsoft Corporation Station-Q. We acknowledge Solliance, a solar energy R&D initiative of ECN, TNO, Holst, TU/e, imec and Forschungszentrum Jülich, and the Dutch province of Noord-Brabant for funding the TEM facility. The work at University of California, Santa Barbara was supported in part by Microsoft Research. We also acknowledge the use of facilities within the National Science Foundation Materials Research and Science and Engineering Center (DMR 11–21053) at the University of California: Santa Barbara and the LeRoy Eyring Center for Solid State Science at Arizona State University.

**Author Contributions**

S.G., D.C., J.L., C.J.P. and E.P.A.M.B. carried out the material synthesis. H.Z. and M.W.A.M. fabricated the devices and performed the transport measurements and data analysis. S.C.B, M.C.C. and R.S. carried out the substrate preparation. D.X. and G.W. fabricated the hard gap devices and contributed to the measurement. S.G. and R.L.M.O.H.V. did the nanowire manipulation for the TEM analysis and transport measurements. M.A.V. performed TEM analysis. B.S. and D.P. J.S.L. contributed to the experiments at University of California Santa Barbara. S.K. prepared the lamellae for TEM analysis. J.S. and D.B. contributed to the hard gap device fabrication. P.J.V.V. gave the support on the MOVPE reactor. E.P.A.M.B, C.J.P. and P.K. provided key suggestions on the experiments. E.P.A.M.B., C.J.P. and L.P.K. supervised the projects. All authors contributed to the writing of the manuscript.

## Extended Data: Epitaxy of Advanced Nanowire Quantum Devices

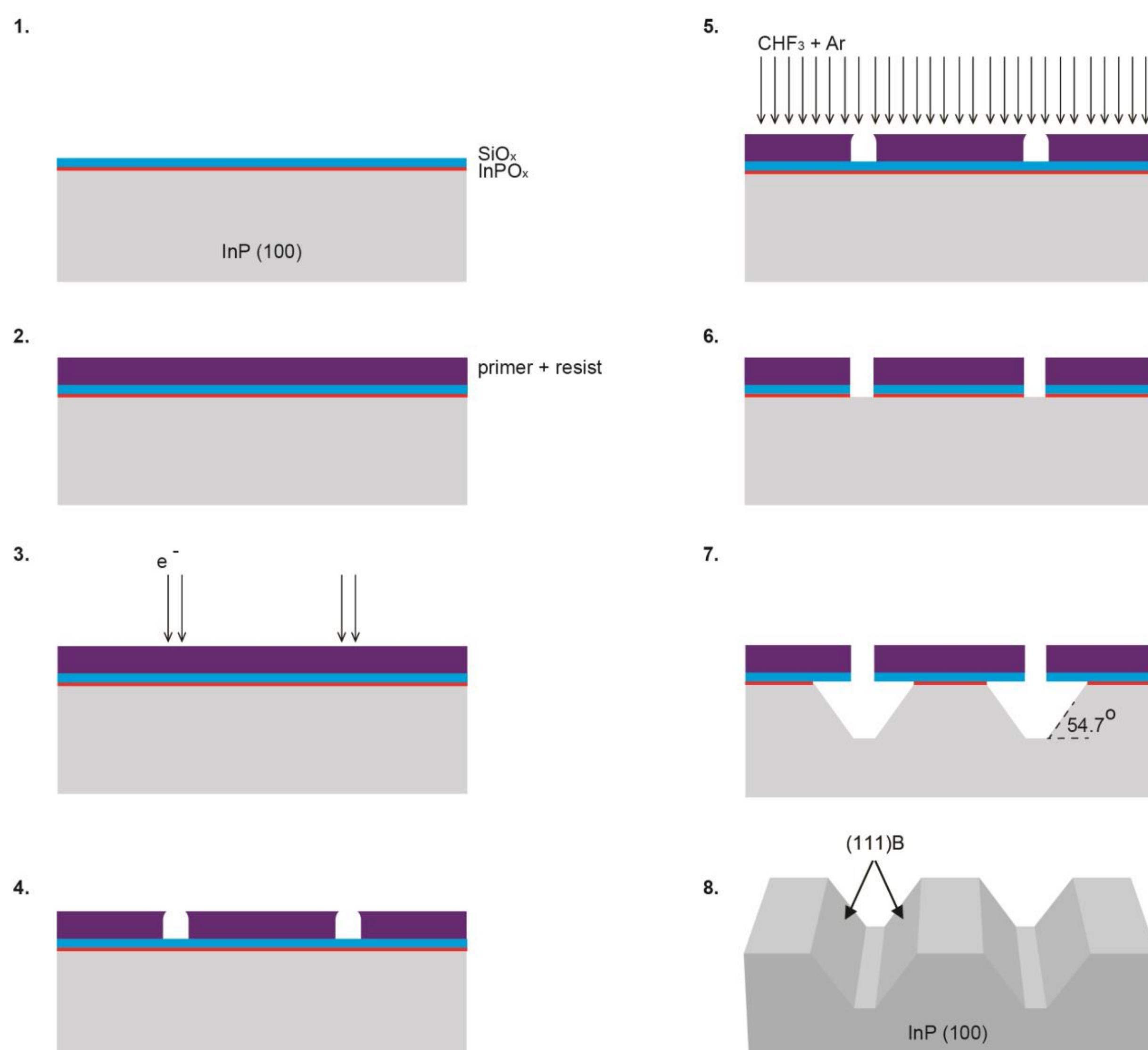

**Extended Data Figure 1: Fabrication of InP substrate with trenches. 1-8**: Schematic illustration of the processing steps. **1,** An out of the box wafer is etched in 7:1 buffered HF; An oxygen plasma step is performed to create a "sacrificial" native oxide layer of 1.9 ± 0.1 nm[29]; A 20 nm $SiO_x$ hard mask is deposited followed by another oxygen plasma treatment. **2-3,** The e-beam primer and resist layer is spun; Rectangular windows of ~ 200 nm are written using e-beam lithography and developed subsequently. **5-6,** The hard mask is etched using reactive ion etching (RIE) with $CHF_3$ and Ar. **7-8,** The wet etch in HCl (37%):$H_3PO_4$ (85%) with 5:1 ratio is performed to expose (111)B facets in InP (100) and the hard mask is removed using 7:1 buffered HF. A detailed fabrication recipe can be found in Supplementary Note 1.1b.

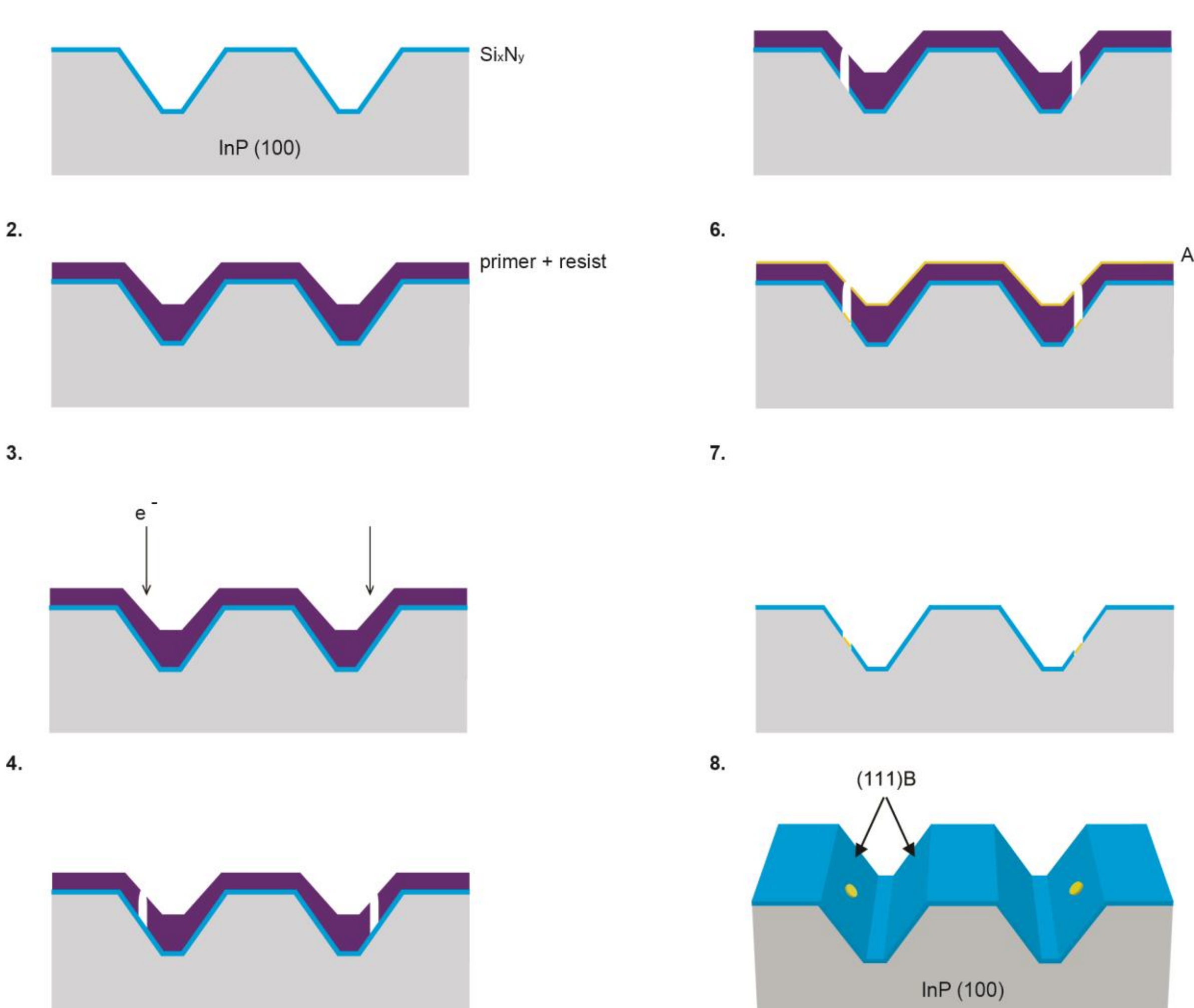

**Extended Data Figure 2: Catalyst deposition. 1-8**: Schematic illustration of the processing steps. **1,** 20 nm $Si_xN_y$ mask is deposited followed by an oxygen plasma treatment. **2-4,** E-beam primer and resist layer is spun (nominal resist thickness needs to be half of the depth of the trenches); Arrays of dots (10-50 nm) are written on inclined (111)B facets using e-beam lithography and the resist is then developed. **5,** Openings in $Si_xN_y$ mask are defined using short 20:1 buffered HF etch. **6-8,** 10 nm of gold is evaporated through the opening in $Si_xN_y$ mask followed by a lift-off. A detailed fabrication recipe can be found in Supplementary Note 1.1c.

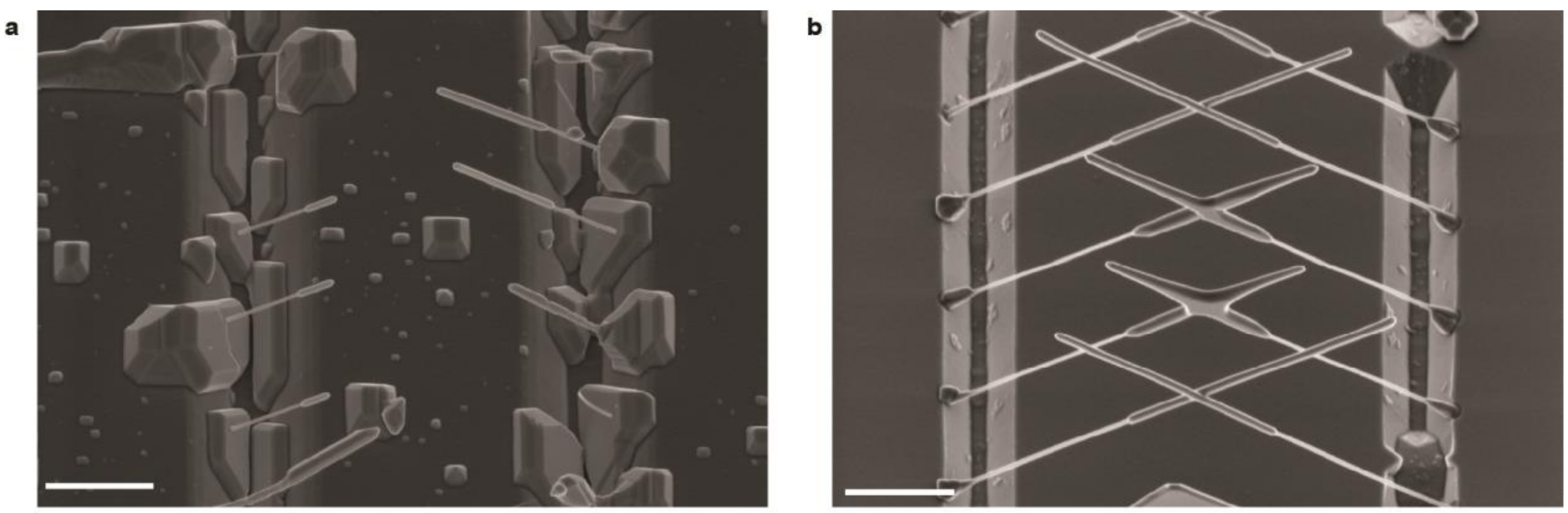

**Extended Data Figure 3: Role of the $Si_xN_y$ mask in InSb NW growth.** A 30°-tilted SEM image of InP-InSb NWs grown on a substrate without (**a**) and with (**b**) $Si_xN_y$ mask. A substantial amount of parasitic thin film growth is observed in **a**. Concave edges of the trenches act as a preferential nucleation site for InSb growth. Thin film InSb growth is in direct competition with InSb NW growth, resulting in short NWs and a very low yield of crossed junctions. By covering the substrate with a $Si_xN_y$ mask, the growth is restricted to areas where the InP substrate is exposed[30]. This, in combination with ~100 times lower molar fractions of TMIn and TMSb used for the growth of wires shown in **b**, eliminates the unwanted InSb layer growth and allows for growth of high-aspect ratio InSb NWs which merge into networks. Both scale bars are 1 µm.

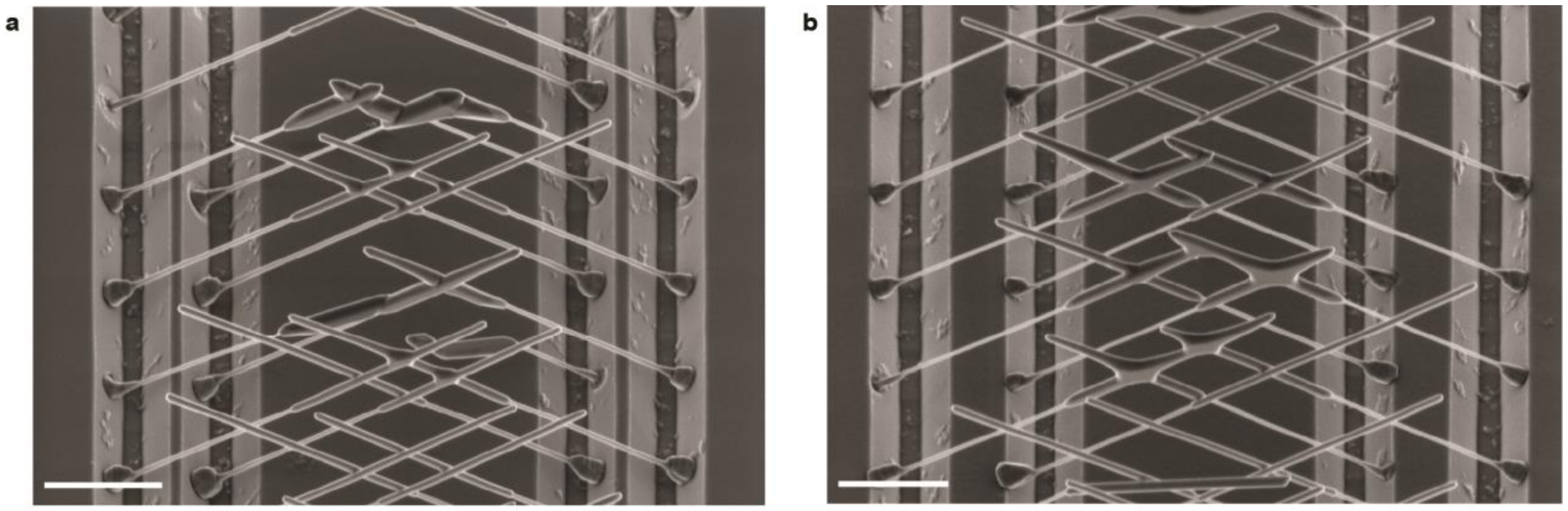

**Extended Data Figure 4: Lithographic control over the trench design layout enables growth of "hashtags" spanning different loop areas. a-b,** InSb NW networks grown on trenches with different spacing between the *left-left* ($L_1$, $L_2$) and *right-right* ($R_1$, $R_2$) trenches, labelled *a* and *b* in Fig. 1a, respectively. Control over the dimensions of the trenches allows us to tune the length of the "hashtag" parallelogram. The scale bar is 1 μm.

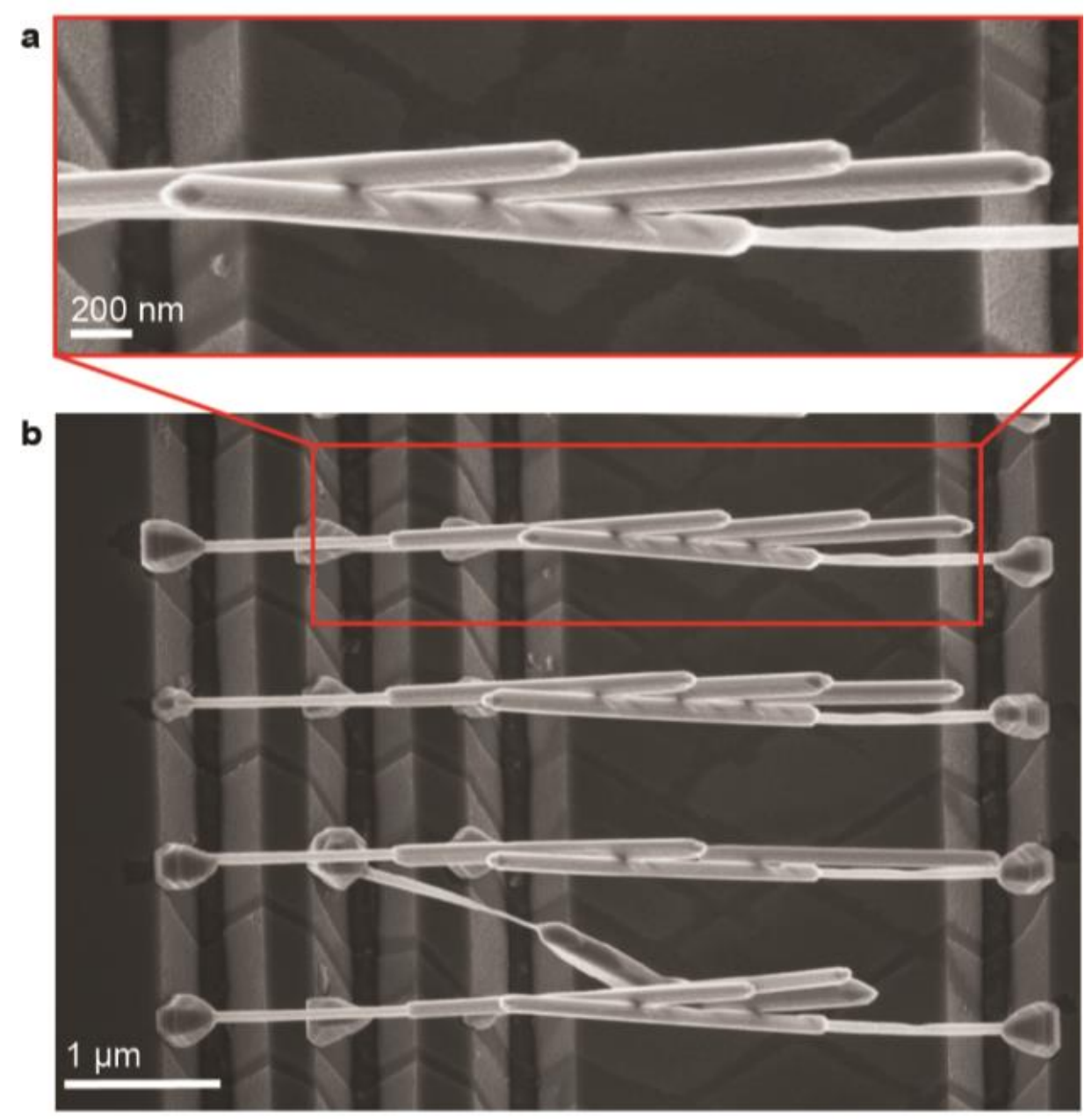

**Extended Data Figure 5: The number of SCIs, *n+1*, is determined by the number of wires, *n*, directly in front of the shadowed nanowire. a,** A high magnification top-view SEM image of the region indicated by a red rectangle in panel **b,** The three NWs facing the Al flux cast shadow on the wire directly behind them, resulting in InSb NWs with 4 SCIs. The shadowing offset is ~200 nm. The NWs bend toward each other due to the e-beam exposure during imaging.

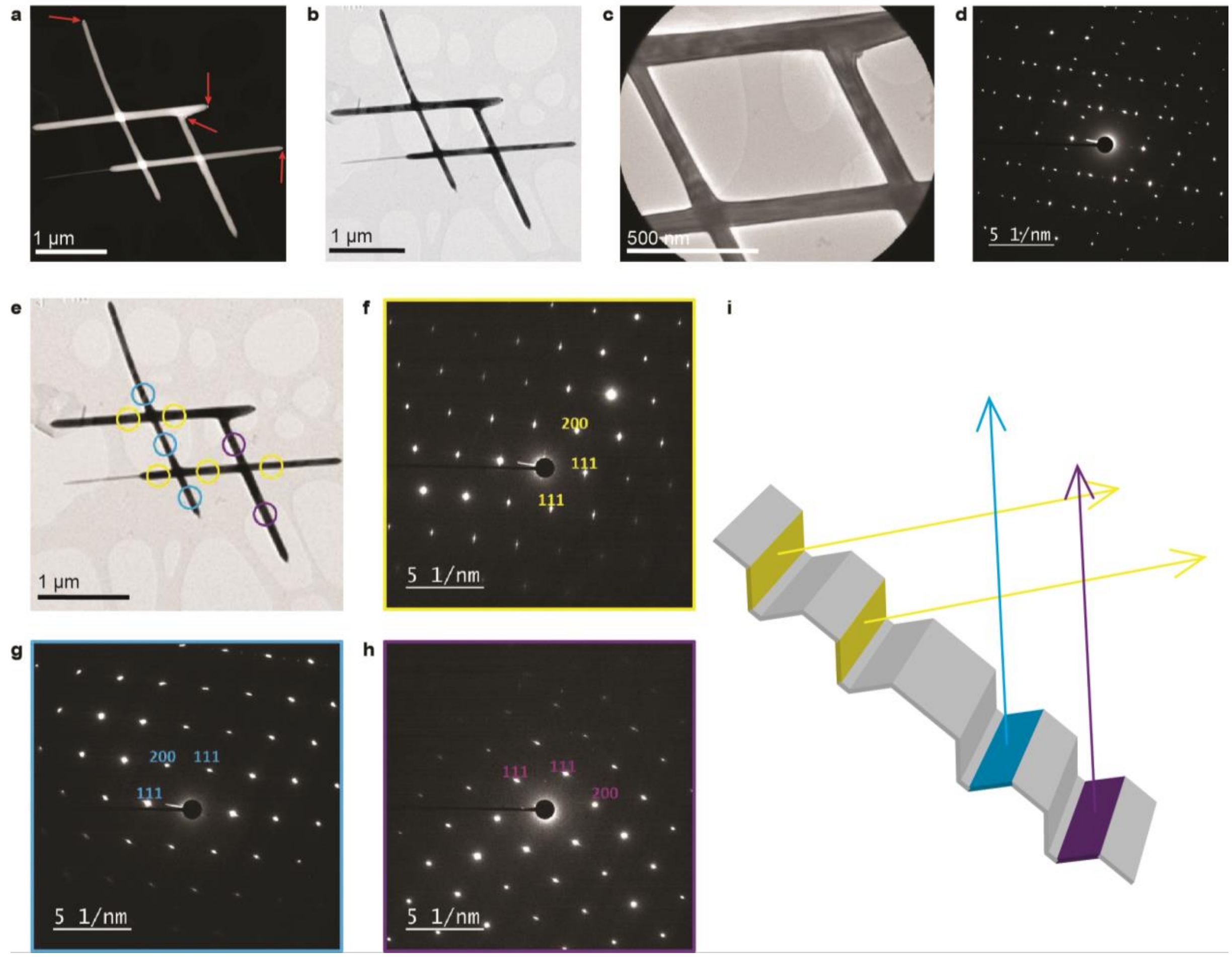

**Extended Data Figure 6: Structural analysis of a "hashtag" taken from the substrate and deposited on a holey carbon film using a micromanipulator in the SEM. a,** High Angle Annular Dark Field (HAADF) Scanning TEM image of the hashtag. The red arrows indicate the positions of the gold catalyst particles. For one wire, the InP stem is present and recognizable. **b,** Corresponding Bright Field (BF) TEM image. **c,** BFTEM image displaying the central part of the hashtag as well as the 1.3 µm aperture inserted for the Selected Area Electron Diffraction (SAED) pattern displayed in **d.** The pattern represents a superposition of three twin-related <110> zone axis patterns. **e,** To reveal the orientation of the individual wires of the "hashtag", SAED patterns for all the wires were acquired, using a smaller SAED aperture diameter of 0.25 µm. Three different <110> zone axis patterns were recorded. The colour coding of the apertures in panel **e** corresponds to the patterns in **f**,**g**,**h**. **i,** schematic representation of the formation of the hashtag presented in the TEM images: The "blue" and "purple" nanowires have two different orientations, related by a $180^0$ rotation around their long axis. Thus, one of the wires has the same orientation as the substrate wafer, while the other one is twin related. The two "yellow" wires have identical orientations that differ from the orientations of the two other wires. Thus, these yellow wires are also twin-related to the substrate, though their rotation axis is different from that of the blue and purple wires.

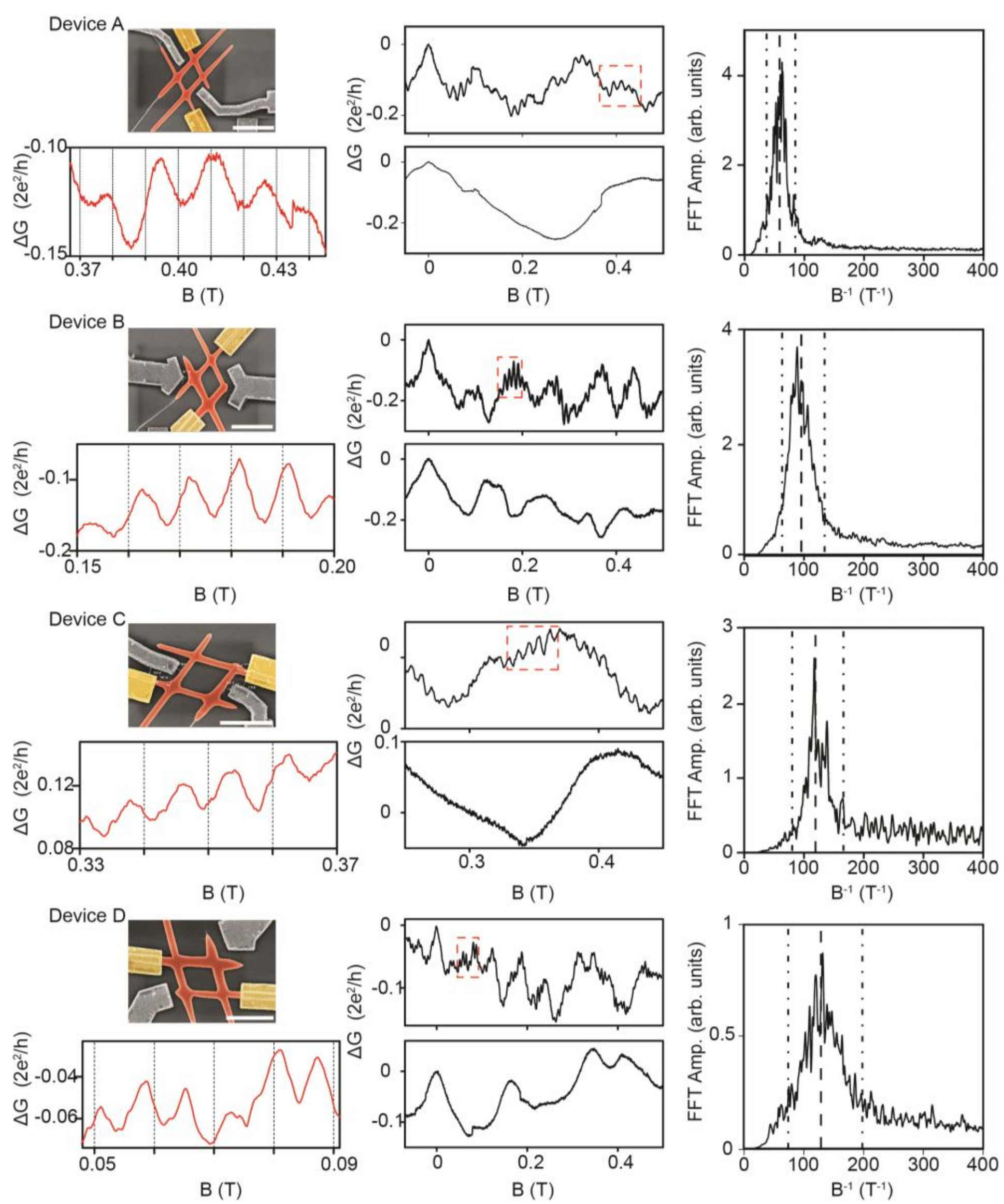

**Extended Data Figure 7: Aharonov-Bohm (AB) oscillations in four devices with different "hashtag" surface areas.** Device A has been studied in detail in the main text. For all devices, left upper panel shows the false-coloured SEM image of the device, middle panels show the conductance measured in the out-of-plane (top) and in-plane (bottom) magnetic field and right panel shows the ensemble averaged FFT spectrum. Only the out-of-plane magnetic field, whose flux penetrates through the "hashtag" loop, gives AB oscillations which indicates that the AB oscillations indeed originate from the coherent interference of electron waves of the two separated conducting nanowire "arms". A magnified view of the AB oscillations (a zoom-in on the region indicated by an orange rectangle in the upper middle panel) is shown in the lower left panel, while the right panel shows averaged FFT spectrum. Plot of the peak frequency, assigned from the averaged FFT spectra, as a function of the measured loop area of the four devices is shown in main text, inset Fig. 3b. Weak-antilocalization peak (WAL) at $B = 0$ T is present for both field directions, and in three (A, B, D) out of four devices, suggesting the strong spin-orbit nature of the InSb NW network. The corresponding back gate voltages of the four devices are: 15 V, 9 V, 12 V and 9 V, respectively. Temperature is 300 mK. The scale bar is 1 μm.

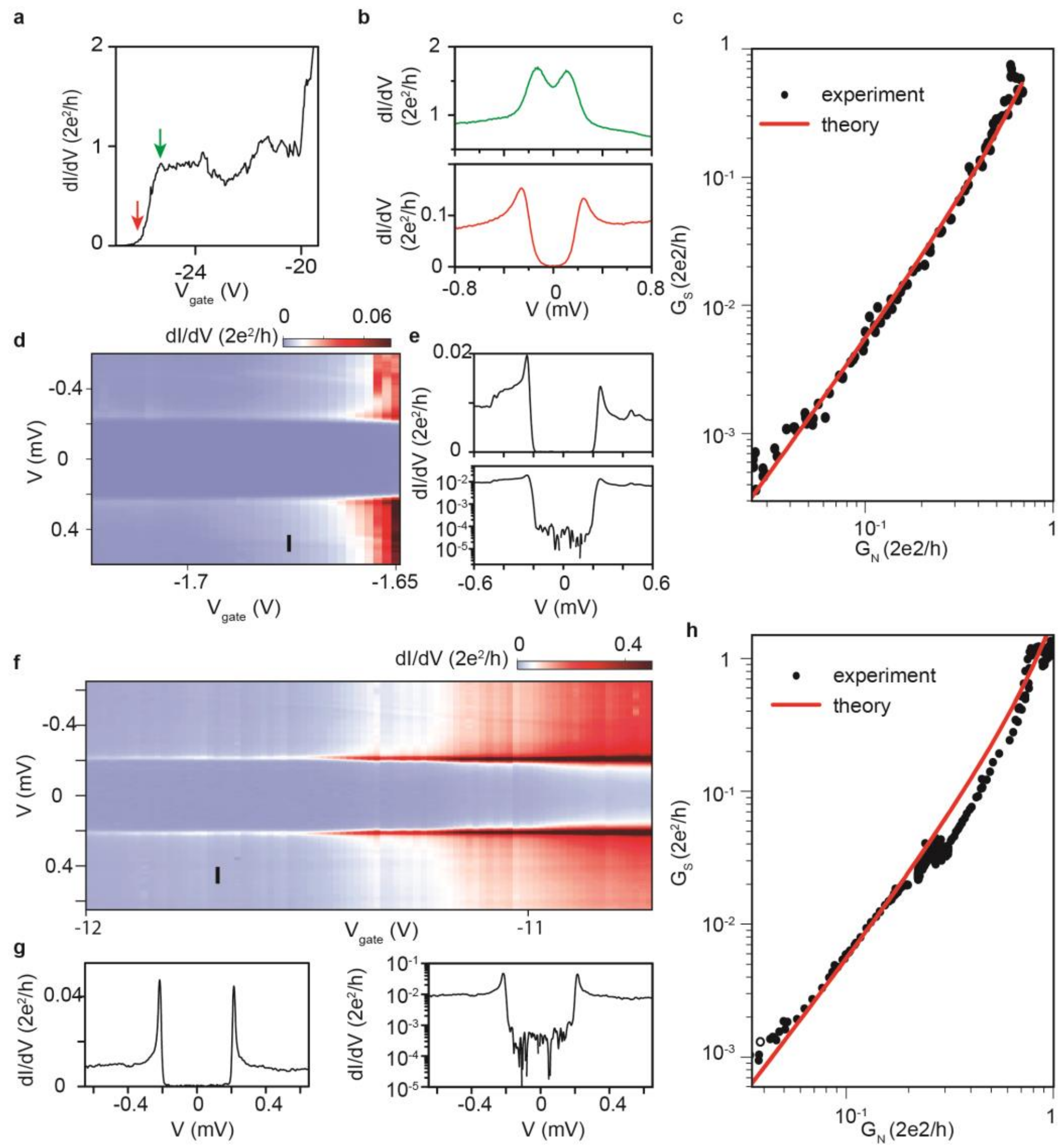

**Extended Data Figure 8: Ballistic transport, Andreev enhancement and hard gap in additional Al-InSb devices. a**, Above-gap (normal carriers) conductance of Device Y as a function of $V_{gate}$. A conductance plateau near the quantized value ($2e^2/h$) can clearly be seen, indicating ballistic transport. **b,** *dI/dV* vs. bias voltage in the open and tunneling regime, resolving strong Andreev enhancement (green) and a hard gap (red), respectively, with $V_{gate}$ indicated by arrows in panel a. The coherence peaks are smeared out due to thermal broadening (temperature ~300 mK for this device). The Andreev enhancement is due to Andreev reflection: an incoming electron reflects as a hole at the NS interface generating a Cooper pair. This process effectively doubles the transported charge from *e* to 2*e*, enhancing the sub-gap conductance. Our enhancement factor reaches 1.7 x $2e^2/h$, indicating the high Al-InSb interface transparency, with transmission larger than 0.96. The small dip in Andreev enhancement near zero bias is due to mode mixing induced by minimal residual disorder[11]. **c,** Sub-gap vs. above-gap conductance of Device Y (black dots), and a theoretical fit (red) based on the Beenakker formula, showing perfect agreement over three orders of magnitude conductance change. **d,** *dI/dV* of Device Z as a function of $V_{gate}$. **e,** A line cut from panel **d** (black bar), plotted in linear (top) and logarithmic scale (bottom). The above-gap/sub-gap ratio is larger than 300. **f,** *dI/dV* of Device M as a function of $V_{gate}$. **g,** A line cut from panel **f** (black bar), plotted on linear (left) and logarithmic scale (right). Device shown in the main text (X), and devices Z and M are measured at ~20 mK.

# Supplementary Information: Epitaxy of Advanced Nanowire Quantum Devices

# 1. Sample fabrication and growth of the networks

## 1.1 Trenches

Fabrication of substrates with trenches is a three-step lithography process:

*First*, Electron-beam lithography (EBL) and metal lift-off are used to deposit alignment markers on a (100) InP substrate.

*Second*, the InP (100) wafer is cleaned in buffered oxide etch ($NH_4F$:HF=7:1) for 5 min and exposed to $O_2$ microwawe plasma to create a thin (~2 nm) sacrificial layer of native oxide on the surface prior to deposition of 50 nm of $SiO_x$ by plasma-enhanced chemical vapor deposition (PECVD). EBL and reactive ion etching (RIE) in $CHF_3$/Ar plasma are used to define rectangular openings in $SiO_x$, whose long edge is aligned with the [0-1-1] direction of the substrate. The alignment of the openings is crucial to achieve trenches with inclined (111)B facets after the subsequent isotropic wet-etch step (HCl:$H_3PO_4$=5:1[1], for 15 s at 1°C).

*Third*, $SiO_x$ is stripped in BHF (5 min) and 20 nm of PECVD $Si_xN_y$ is deposited on the substrate to prevent the parasitic InSb thin film growth which competes with nanowire growth[2]. EBL step followed by a short (40 sec) buffered oxide etch ($NH_4F$:HF=20:1 + surfactant (Triton)) is used to define openings in the $Si_xN_y$ mask. Metal evaporation (8 nm of Au) and lift-off are used to position Au catalysts (10-50 nm in size) in the openings in the $Si_xN_y$. Detailed description of processing steps is listed below.

**a) Markers**

***Substrate Cleaning***

InP (100) wafer is cleaned with buffered oxide etch ($NH_4F$:HF=7:1) (5 minutes), rinsed with $H_2O$, IPA (10 minutes)

***Fabrication of markers***

Spin resist AR-P 6200.13 at 6000 rpm, bake at 150 °C for 3 minutes;

Write marker patterns using e-beam lithography (dose 300 μC/cm$^2$);

Developing in AR 600-546 for 1:30 minutes in ultrasonic agitation;

Ultrasonic rinse in IPA for 30 seconds, blow dry;

Evaporation of 80 nm Au;

Lift-off in PRS3000 at 88 °C for 2 hours;

Rinse in warm ( >50 °C) $H_2O$;

Rinse in IPA for 1 minute, blow dry

**b) Trenches**

***Substrate Cleaning***

InP (100) wafer with markers is cleaned with buffered oxide etch ($NH_4F$:HF=7:1) (5 minutes), rinsed with $H_2O$, IPA (10 minutes)

***Hard mask (1. in Extended Data Fig. 1)***

Sacrificial layer deposition microwave oxygen plasma (10 minutes, 200 mL/min, power 100 watts, PVA Tepla 300);

PECVD 20 nm $SiO_X$ deposition (300 °C, Oxford Instruments PlasmaLab 80 Plus);

Oxygen plasma (60 seconds, power 40 watts).

***Fabrication of trenches (2 – 8 in Extended Data Fig. 1)***

Spin primer (sticking layer) AR 300-80 at 2000 rpm, bake at 180 °C for 2 minutes

Spin resist AR-P 6200.13 at 6000 rpm, bake at 150 °C for 3 minutes;

Write trench patterns using e-beam lithography (dose 350 $\mu C/cm^2$);

Developing in AR 600-546 for 1:30 minutes in ultrasonic agitation

Ultrasonic rinse in IPA for 30 seconds, blow dry;

RIE (reactive ion etch) mask (23 W, 50 sccm $CHF_3$, 2 sccm Ar, Leybold Hereaus, 12 minutes);

Wet etch in HCl (37%) :$H_3PO_4$ (85%) ratio 5:1 (15 seconds, 1 °C);

Strip the resist in PRS3000 at 88 °C for 20 minutes;

Removing hard mask in buffered oxide etch ($NH_4F$:HF=7:1) (5 min).

**c) Catalyst deposition**

***Deposition of the mask (1. in Extended Data Fig. 2)***

PECVD 20 nm $Si_XN_Y$ deposition (300 °C, Oxford Instrumentals PlasmaLab 80 Plus);

Oxygen plasma (60 seconds, power 40 watts)

***Dots formation (2 – 8 in Extended Data Fig. 2)***

Spin primer (sticking layer) AR 300-80 at 2000 rpm, bake at 180 °C for 2 minutes;

Spin resist AR-P 6200.04 at 4000 rpm, bake at 150 °C for 3 minutes;

Write dot patterns using e-beam lithography (dose 700-800 $\mu C/cm^2$);

Developing in AR 600-546 for 1:30 minutes in ultrasonic agitation;

Ultrasonic rinse in IPA for 30 seconds, blow dry;

Opening the holes in $Si_XN_Y$ mask with buffered oxide etch ($NH_4F$:HF=20:1) + 5 drops of surfactant Triton X-100 (40-60 seconds);

Rinse with $H_2O$, IPA (10 minutes);

Evaporation of 10 nm Au;

Lift-off in PRS3000 at 88 °C for 2 hours;

Rinse in warm ( >50 °C) $H_2O$;

Rinse in IPA for 1 minute, blow dry.

## 1.2 Nanowire Growth

To remove organic residues from the wafer caused by the photoresist layer, substrates were descumed in $O_2$ plasma (10 min, 55 sccm $O_2$, 300 W plasma power) prior to loading into an horizontal Aixtron 200 metal-organic vapor phase epitaxy (MOVPE) reactor with infrared lamp heating. InP NWs, which act as the mediator for InSb NW growth, were grown at 450°C for 19 min using tri-methyl-indium (TMI), phosphine ($PH_3$) and HCl (1%) with precursor molar fractions $X_i$ (TMI)= $7.6 \times 10^{-6}$ and $X_i$ (PH3) = $9 \times 10^{-3}$ and $X_i$ (HCl)= $8.3 \times 10^{-6}$. HCl was used to suppress unwanted sidewall growth. InSb nanowires were grown at 495 °C using tri-methyl-indium (TMI) and tri-methyl-antimony (TMSb) with precursor molar fractions $X_i$ (TMI)= $2.8 \times 10^{-7}$ and $X_i$ (TMSb)= $5.1 \times 10^{-5}$, for 35 minutes. For both processes, the reactor pressure was 50 mbar, total flow 6000 sccm and $H_2$ was used as a carrier gas.

## 1.3 Growth of Superconducting Aluminium Islands

NW networks are transferred *ex-situ* to a molecular beam epitaxy (MBE) chamber where the atomic hydrogen clean (20 min under continuous rotation, 380°C, $5x10^{-6}$ Torr $H_2$ pressure) is first done to remove the native oxide from the InSb NW surface[3]. Subsequently, samples were cooled down to ~120 K by active liquid nitrogen cooling. Careful alignment of NWs relative to the Al source is important for well-controlled shadowing of the NWs. Samples are aligned such that the Al flux is parallel to the long edge of the trenches, as illustrated in Fig. 2a. Al cell temperature was 1085 °C, resulting in growth rate of ~2 Å/min. Immediately after growth, samples were transferred *in-situ* to an MBE chamber equipped with ultra-high purity $O_2$ source where they were dosed with $\sim 10^{-5}$ Torr of $O_2$ for 15 mins. This step is important because a so-formed self-terminating oxide layer will “freeze-in” the Al film, preventing it from diffusing and forming Stranski-Krastanov Al islands, while the sample is being heated up to room-temperature in UHV, prior to unloading from the MBE chamber.

# 2. Device fabrication and transport analysis

## 2.1 Aharonov-Bohm devices

**Device fabrication recipe:**

1. Transfer "hashtag" nanowires onto a p-doped Si substrate covered by 285 nm $SiO_2$ layer, serving as a back gate dielectric.
2. Spinning bi-layer PMMA: first PMMA 495K A6 at 3000rpm spinning rate, bake at 175 degrees for 10 minutes. Then PMMA 950K A2 at 2000rpm, bake at 175 degrees for 10 minutes.
3. Write designed contacts and side gates patterns with e-beam.
4. Develop in developer (MIBK:IPA=1:3) for 1 minutes, clean in IPA for 1 minutes, air-gun blow dry.
5. Descum, Oxygen plasma 1 minute with power 100 W, pressure 1.95 mbar. (with Faraday cage in to screen the plasma).
6. Sulfur passivation: dip the chip in ammonium sulfide solution (3 ml $(NH_4)_2S$ mixed with 290 mg sulfur powder, then diluted with DI-water at a volume ratio of 1:200) at 60 degrees for 30 minutes. Then, rinse the chip in DI water and transfer to an evaporator.
7. Helium milling for 30 seconds with a Kauffman ion source. Then continue to evaporate Au/Cr (200 nm/10 nm).
8. Lift off in acetone.

**Measurement and Analysis:**

All the four AB devices were measured in a He-3 fridge with based temperature ~300 mK. During the measurement, the side gates (grey in Fig. 3a left inset) were kept grounded, and the global back gate is used to turn on the conducting channels in the “hashtag arms”. The back gate voltage is 13.35 V for the measurement in Fig. 3a.

**Ensemble average of FFT**: The FFT spectrum shown in Figure 3b is an ensemble average of the absolute values of 25 individual FFT spectra[4]. The individual FFTs were calculated from the corresponding magnetoconductance traces (including the one in Fig. 3a), which were measured successively with gate voltage values between 13.3 V and 13.7 V (resulting in conductance values between 0.7 and 0.9 x $2e^2/h$). A smooth background is subtracted from the original magnetoconductance curves before the FFT is calculated.

**Estimation of “hashtag” loop area:** The estimation is based on the SEM images of the device. We took the middle of the wire as the loop boundary to estimate the area, while the error bar of the area is estimated based on the accuracy of the nanowire length we measured from SEM images.

**Estimation of phase coherence length**: The amplitude of the AB oscillations is determined by integrating the obtained Fourier spectrum over the frequency range corresponding to the expected $h/e$ peak. This amplitude decays as $\exp(-L/L_\phi(T)$, where $L$ is the relevant device length, and $L_\phi$ is the phase coherence length. For a quasiballistic system coupled to a thermal bath, we have $L_\phi \propto T^{-1}$ giving an exponential dependence of the AB amplitude on temperature[5]. The slope of the fitted curve then gives the ratio between $L$ and $L_\phi$ at 1 K. This allows us estimate the phase coherence length if the relevant device length scale is known.

## 2.2 Hard gap devices

InSb NW with two shadowed aluminium islands is contacted by Au/Cr. Argon plasma etching was used to remove the aluminium film prior to evaporation of normal contacts. One normal contact is deposited right next to the shadowed region to replace one aluminium island. The second normal contact is on the other end of the nanowire, sufficiently apart from the shadowed region not to affect the superconducting properties in its vicinity, serving as a current drain for the superconducting contact.

**Device fabrication recipe:**

1. Transfer InSb-Al nanowires onto a p-doped Si substrate covered by 285 nm $SiO_2$ layer, serving as a back gate dielectric.
2. Spinning PMMA 950K A6 at 4000 rpm, leave the chip in a vacuum chamber pumped with a turbo for overnight.
3. Write designed electrode contact patterns with e-beam, beam dosage: 2300, 1900 and 1800 $\mu C/cm^2$ for fine, coarse and bonding pads pattern, respectively.
4. Development: (MIBK:IPA=1:3) for 1 minute, IPA for 1 minute, blow dry.
5. Ar plasma etch for 4 minutes (with Ar pressure 3 mTorr, 100 Watts) to etch away Al, $AlO_x$, InSb surface oxide and part of the InSb nanowires. To prevent the PMMA from burning due to long plasma etch, one can perform short plasma etch (*e.g.* 20 s) for 12 times with 40s break between each etch to let the chip to cool down.
6. Evaporate normal contacts, Au/Cr (100 nm/10 nm).
7. Lift off in acetone.

**Measurement and Analysis**

Device X (main text), Z, M were measured in a dilution refrigerator with a base temperature ~20 mK, while Device Y was measured in a He-3 fridge with a base temperature ~300 mK.

**Sub-gap vs above-gap conductance fitting in Al-InSb devices**: we assume there is a single transmitting channel in the shadow region with transmission $T$. The above-gap conductance is conductance of normal carriers: $G_N = 2e^2/h \times T$, while sub-gap conductance, based on Beenakker's formula[6], is: $G_S = 2e^2/h \times 2T^2/(2-T)^2$. Thus $G_s$ can be plotted as a function of $G_N$ as shown in Fig.4c (red line). For the experimental data, at each gate voltage, we get the above-gap conductance by averaging the conductance at bias ($V$) much larger than the gap ($\Delta$), while the sub-gap conductance is obtained by averaging a small bias window at zero bias.

**Contact transparency estimation based on Andreev enahancement**: in extented data Fig. 10b, we get Andreev enhancement for subgap conductance reaching 1.7 x $2e^2/h$. Based on Beenakker's formula, setting this value equal to $G_S$, we can extract a transparency $T \sim 0.96$.

**References:**

1. Adachi, S. & Kawaguchi, H. Chemical etching characteristics of (001) InP. *Journal of The Electrochemical Society* **128**, 1342 (1981).

2. Dalacu, D. *et al.* Selective-area vapour–liquid–solid growth of InP nanowires. *Nanotechnology* **20**, 395602 (2009).

3. Webb, J. L. *et al.* Electrical and Surface Properties of InAs/InSb Nanowires Cleaned by Atomic Hydrogen. *Nano Letters* **15**, 4865 (2015).

4. Meijer, F. E. *et al*. Statistical significance of the fine structure in the frequency spectrum of Aharonov-Bohm conductance oscillations. *Physical Review B* **69**, 035308 (2004).

5. Stern, A. *et al.* Phase uncertainty and loss of interference: A general picture. *Physical Review A* **41**, 3436 (1990).

6. Beenakker, C. W. J. Quantum transport in semiconductor-superconductor microjunctions. *Physical Review B* **46**, 12841 (1992).